\documentclass[preprint]{aastex}

\shorttitle{Quick-MESS}
\shortauthors{Bonavita et al.}

\begin{document}

\title{Quick-MESS: \\
A fast statistical tool for Exoplanet Imaging Surveys}

\author{Mariangela Bonavita, Ernst J.W. de Mooij, Ray Jayawardhana}
\affil{Department of Astronomy and Astrophysics - University of Toronto, 50 St. George Street M5S 3H4 Toronto ON Canada}

\begin{abstract}
Several tools have been developed in the past few years for the statistical analysis of the exoplanet search surveys, mostly using a combination of Monte-Carlo simulations or a Bayesian approach.
Here we present the Quick-MESS, a grid-based, non-Monte Carlo tool aimed to perform statistical analyses on results from and help with the planning of direct imaging surveys. Quick-MESS uses the (expected) contrast curves for direct imaging surveys to assess for each target the probability that a planet of a given mass and semi-major axis can be detected. By using a grid-based approach Quick-MESS is typically more than an order of magnitude faster than tools based on Monte-Carlo sampling of the planet distribution. In addition, Quick-MESS is extremely flexible, enabling the study of a large range of parameter space for the mass and semi-major axes distributions without the need of re-simulating the planet distribution. 
In order to show examples of the capabilities of the Quick-MESS, we present the analysis of the Gemini Deep Planet Survey and the predictions for upcoming surveys with extreme-AO instruments.
\end{abstract}

\keywords{Stars, Extrasolar Planets, Data Analysis and Techniques}

\section{Introduction}
More than a decade of extensive searches have led to a sample of over 850 confirmed exoplanets\footnote{http://exoplanets.eu, January 4, 2013} and thousands more planet candidates \citep{2012arXiv1202.5852B}, almost entirely identified through indirect detection techniques. Such a large number of discoveries allows accurate statistical analyses to address questions related to the distribution of their properties, such as the mass, orbital period and eccentricity \citep{2003ApJ...598.1350L,2008PASP..120..531C} as well as the relevance of the host star characteristics (mass, metallicity) on the final frequency and distribution of planetary systems \citep[see][]{2005ApJ...622.1102F, 2004A&A...415.1153S, 2007ApJ...670..833J}. Since the most successful techniques (radial velocity and transit) have focused on the inner ($\le 5 AU$) environment of main sequence solar-type stars, most of the available information on the frequency of planets concerns this phase space. Radial velocity (RV) surveys report that $\sim$19\% of nearby Sun-like stars harbor planets within 20 AU \citep{2008PASP..120..531C}, and recent \textit{Kepler Space Telescope} results indicate the frequency of stars with planets to be $\sim$30\% \citep{2011ApJ...736...19B}.

Since both the transit and the radial velocity techniques are biased towards planets in relatively close orbits, orbital separations larger than $\sim$5 AU are currently not well sampled. Direct imaging surveys, which are typically more sensitive to planets at larger orbital separations can fill this gap. However, to predict the expected planet fractions, the distributions derived from the radial velocity and transit surveys need to be extrapolated.  Several tools have been developed in the past few years for the statistical analysis of direct imaging (DI) planet search surveys \citep[see e.g.][]{2010A&A...509A..52C,2008ApJ...689L.153L,2008ApJ...674..466N, 2012A&A...537A..67B}, which try to assess at what confidence planet distributions obtained from RV surveys can be extrapolated to estimate the planet distributions to the orbital separations where DI is more sensitive.

\cite{2012A&A...537A..67B} (hereafter Paper 1) describes the Multi-purpose Exoplanet Simulation System (MESS) an IDL simulation code specifically designed to perform statistical analysis of direct imaging surveys\footnote{available for download at www.messthecode.com}. 

The MESS combines the properties of the target stars (mass, luminosity, distance, age, etc.) with the assumptions on the planet parameter distributions (PPD), to generate a synthetic planet population (SPP) and the measured/estimated detection limits of the instrument under scrutiny, to estimate the probability of detecting a companion. This is then used to validate the original assumption on the PPD, in light of the observed or predicted detection limits. One can then explore different sets of PPD, the goal being to constrain those that are compatible with the observations. In case of planned observations, this information can be used to determine which kind of constraints on the PPD the new instrument will be able to provide, and, given a certain PPD, what the expected detection rate would be. 

In this paper we present a novel way to perform statistical analyses of DI surveys, called Quick-MESS, where the standard Monte Carlo approach is replaced by grid-based sampling of the orbital parameters, which leads to a substantial reduction in the required computational time. The main features of the code are described in Sec.~\ref{sec:qmess}. In Sec.~\ref{sec:app} we present the two main applications of the code, and in Sec.~\ref{sec:summary} we provide a summary.

\section{Quick MESS}
\label{sec:qmess}

The Quick-MESS (hereafter QMESS) code presented here does not use a Monte-Carlo approach as used by most of the statistical tools use so far for the analysis of exoplanet surveys results \citep[see e.g.][]{2012A&A...537A..67B, 2010A&A...509A..52C, 2008ApJ...689L.153L}. Instead it takes a grid-based approach consisting of several steps that allow the determination of the probability of detecting a planet in the considered parameter space, given a set of assumptions on the distribution of the planet orbital parameter and the mass-luminosity function. The main steps of the code, explained in details in sections \ref{sec:prjprob} and \ref{sec:detprob}, can be summarized as follows:

\begin{enumerate}
\item Evaluate the distribution of planets with a certain normalised separation, $s$, as a function of the orbital parameters (as discussed in Sec.\ref{sec:prjprob}) and integrate and normalise it to get $f(s,e)$: the distribution of planets as a function of the eccentricity $e$  and the normalized separation $s$, where {\it s} is defined as {\it s}=R/a with $R$ the physical separation and $a$ the planet's semi-major axis.
\item Multiply $f(s,e)$ for the eccentricity distribution $f(e)$ 
\item Use the planetary evolutionary models \citep[e.g][]{2003A&A...402..701B} to estimate, for each target in the studied sample, 
the minimum detectable planetary mass $M_{lim}$ as a function of the projected separation ($\rho$), given the contrast limits of the instrument.
\item For each value of semi-major axis on an uniform grid, use the distance of the star to convert the normalised separation $s$ into the physical projected separation $\rho$, thus obtaining $f(\rho, e)$ from $f(s, e)$.
\item For each value of the planetary mass ($M_p$) over an uniform grid find the range of projected separations where $M_P > M_{lim}$ and integrate $f(\rho, e)$ over this interval to obtain the distribution of detectable planets with a given value of mass and semi-major axis $f(Mp, a)$ 
\item Finally multiply $f(M_p, a)$ with the required mass and semi-major axis distributions ($f(M_p)$ and $f(a)$, respectively) to obtain the detection probability map $g(M_p, a)= f(M_p, a) f(M_p) f(a)$
\item In the case of a survey, integrate $F(M_p, a)$ over the whole range of masses and semi-major axis, in order to obtain $P(M_{min} \leq M_p \leq M_{max}, a_{min} \leq a \leq a_{max})$ the probability of detecting a planet in the considered parameter space.
\end{enumerate}

\subsection{Evaluation of the projection probability}
\label{sec:prjprob}
The distribution of planets that, given a certain combination of orbital parameters, can be found at a certain position on their projected orbit is assessed by calculating the orbit of the planet in normalized separation $s\left(\phi\right)  = R\left(\phi\right)/a$, (where $a$ is the semi-major axis of the orbit, and $R$ is the radius vector that, together with the true anomaly $\nu$ gives the polar coordinates of the planet on the orbit) as a function of the orbital phase $\phi$, over a finely sample grid of $0\leq \phi \leq 1$ and for a range of eccentricities $0 \leq e < 1 $.
For a given instant in time $t$, and being $T_0$ the time of periastron passage and $p$ the orbital period of the planet, the orbital phase $\phi$ is defined as $\phi=\frac{t-T_0}{p}$, which is used to calculate 
the mean anomaly $M$ as in Eq.~\ref{eqn:m_anom}.
The eccentric anomaly $E$ is then calculated using Eq.~\ref{eqn:ecc_an} where $E_0$ and $M_0$ are 
introduced as approximations of $E$ and $M$ and the whole calculation is repeated until the final result is stable \citep[see][]{1978GAM....15.....H}.

\noindent Finally the true anomaly $\nu$ is evaluated using Eq.~\ref{eqn:tr_anom}

\begin{eqnarray}
M   & = & \left(\frac{t-T_0}{p}\right)2\pi  =  2\pi\phi  =  E - e \sin E \label{eqn:m_anom}\\
E_0 & = & M + e \sin M + \frac{e^2}{2} \sin 2M \nonumber \\
M_0 & = & E_0 - e\sin E_0 \nonumber\\
E   & = & E_0 + \left(M-M_0)\right)/\left(1-e\cos E_0\right)\label{eqn:ecc_an}\\
\tan{\nu/2} & = & \sqrt{ \left(1+e\right)/\left(1-e\right)} \tan{E/2} \label{eqn:tr_anom}
\end{eqnarray}

\noindent Given $\nu$, we can define the normalised separation $s$: 
\begin{equation}
s= R/a = \cos \left(\nu + \omega\right) \sec \left(\theta - \Omega \right)\label{eqn:rho_fin}
\end{equation}

\noindent In Eq.~\ref{eqn:rho_fin} $\omega$ is the argument of periastron, $\Omega$ is the longitude of node and the angle $\theta$ together with the projected separation $\rho$ is used to define the coordinates of the planet on the projected orbit. The angle $(\theta - \Omega)$ is defined as:
\begin{equation}
\tan \left( \theta - \Omega \right) = \tan \left( \nu + \omega \right) \cos i \label{eqn:phi}
\end{equation}

\noindent where $i$ is the inclination of the orbit.

Using Eqn.~\ref{eqn:rho_fin} f(s, e, $\omega$, $i$), the distribution of planets found at a given separation, given $e$, $\omega$, and $i$, can be calculated. Since $\omega$ and $i$ are assumed to be uniformly distributed, they can be integrated over in order to give f(s, e): the distribution of planets with a given normalised separation and eccentricity. \footnote{Note that the dependence on the longitude of the node, $\Omega$, is removed using Eqn.~\ref{eqn:phi}}.

\begin{equation}
f(s,e)  =  \int_0^{360} \int_0^1 {f(s,e,\omega,\cos i)  d\omega d\cos i} \label{eqn:pse} 
\end{equation} 

For QMESS, this is done on a fine grid in $\omega$ and $\cos i$, with $0 \leq \omega < 360$ and $0 \leq \cos i \leq 1$ and a step sizes of $\Delta\omega$=1$^{\circ}$ and  $\Delta \cos i$=0.01 respectively.  This map is computationally intensive but only needs to be calculated once. The map used for QMESS, consists of 1000 steps in $e$ and 2000 steps in $s$, is shown in Fig.~1. Note that f(s,e) is assumed to be uniformly distributed in $e$.

Note that altough $f(s,e)$ is, strictly speaking, a probability density function (representing the fraction of planets with a certain combination of eccentricity $e$ and normalised separation $s$.), we will for simplicity from now on refer to it as {\it projection probability}.

\begin{figure}[htbp]
\label{fig:orbit}
\epsscale{0.6}
\plotone{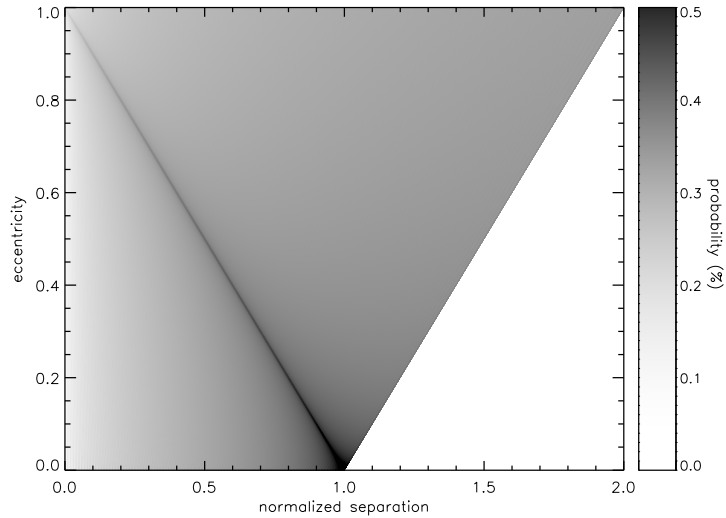}
\caption{\linespread{1} \footnotesize Projection probability, as a function of the eccentricity and normalised separation. Note that the angular dependencies ($\omega, \Omega, i$) have been integrated out (see text)}
\end{figure}

\subsection{Calculating the detection probability}
\label{sec:detprob}
The projection probability f(s,e) obtained in Sec.~\ref{sec:prjprob} is necessary to evaluate the probability of finding a planet of mass $M_p$ and separation $a$, which is the final goal of QMESS. 
Together with f(s,e) a set of distributions for the planet eccentricity ($f(e)=\frac{dN}{de}$), mass ($f(M_p)=\frac{dN}{dM_p}$) and semi-major axis ($f(a)=\frac{dN}{da}$) is also needed. We assume that f(e), f(M$_p$) and f(a) are independent. The implication of this assumption have been discussed by e.g. \cite{2008PASP..120..531C} and \cite{2010ApJ...725.2166H}. In order to account for a non-uniform eccentricity distribution, f(s,e) is multiplied by f(e), although this slightly decreases the flexibility later on, it significantly reduces the memory usage.

The next step consists in converting the instrument detection limit (expressed in minimum planet/star contrast detectable in the chosen band for a given target, as a function of the projected separation $\rho$)
into a minimum planet mass ($M_{lim}$) vs projected separation limit. 
To do this, a set of mass-luminosity models are used \citep[e.g.][]{2003A&A...402..701B,2003ApJ...596..587B}, assuming that the planet and the star are 
coeval. The uncertainties introduced by this approach are discussed in detail in Paper 1. 

Subsequently, the code generates an uniform grid of masses, $M_{min} < M_p < M_{max}$, and semi-major axes, $a_{min} < a < a_{max} $, and evaluates the distribution of detectable planets $f(M_p, a)$.
First, for each semi-major axis we use the distance $d$ of the star to evaluate the projected separation $\rho =\frac{s a}{d}$ and obtain $f(\rho, e)$ for each value of a on the grid.
Then for each value of $M_p$ the values of $\rho_{min}$ and $\rho_{max}$ are evaluated from the detection limits, such as $M_p \geq M_{lim}$  
for $\rho_{min} \leq \rho \leq \rho_{max}$. Note that QMESS assumes that the contrast curve is smooth and has only one (local) minimum, i.e. that the contrast curve crosses a given level of M$_p$ no more than two times. This assumption should hold for most cases, although a local bias from a (bright) nearby companion could affect the results.

 $f(M_p, a)$, defined as the distribution of detectable planets as a function of $M_p$ and $a$, is then calculated as:

\begin{equation} 
f(M_p, a)=\int_0^1{\int^{\rho_{max}\left(M_p, a\right)}_{\rho_{min}\left(M_p, a\right)}}{f\left(\rho, e\right)d\rho de} 
\end{equation}

\noindent The limits on $\rho$ ($\rho_{min}$ and $\rho_{max}$) are defined by the minimum and maximum separation at which a planet is detectable given the contrast curve. 

$f(M_p,a)$ is then stored for each target. 

Note that $M_p$ and $a$ are uniformly distributed in $f(M_p,a)$, although $f\left(M_p, a\right)$ itself is not normalised.
The (expected) distribution for semi-major axes $f(a)$ and planet mass $f(M_p)$, all normalised, are then folded into the $f(M_p,a)$ to provide $g(M_p,a)= f(M_p,a) ( f(a)  f(M_p)$, a new distribution of detectable planets, now taking into account the observed/predicted distribution of planets. 

In a similar way as for $f(s,e)$, we will refer to $f(M_p,a)$ and $g(M_p,a)$ as {\it detection probability} or {\it detection map} although these are also defined as probability density functions. They in fact represent the fraction of detectable planets with a given mass, $M_p$, and semi-major axis, $a$, assuming $f(a)$ and $f(M_p)$ as the distributions of those parameters.

This approach assumes that the distributions of mass, eccentricity and semi-major axes are not correlated. The implication of this assumption have been widely addressed by previous works, including \cite{2008PASP..120..531C} and \cite{2010ApJ...725.2166H}.

Finally $g(M_p,a)$ is integrated over the considered range of mass and semi-major axis to obtain 
the probability of detecting a planet with $M_{min} \leq M_p \leq M_{max}$ and $ a_{min} \leq a \leq a_{max}$ as defined by Eq.~\ref{eqn:probam}

\begin{equation}
P(M_{min} \leq M_p \leq M_{max}, a_{min} \leq a \leq a_{max}) = \int_{M_{min}}^{M_{max}} \int_{a_{min}}^{a_{max}} g(M_p, a) d M_p da 
\label{eqn:probam}
\end{equation}

$P$ can be also used to evaluate the upper/lower limits on the frequency of planets in the range of mass and semi-major axis explored by the 
analysed survey as a function of the assumptions made on the mass and semi-major axis distributions \citep[see e.g.][]{2007ApJ...670.1367L, 2012A&A...544A...9V}.

\subsection{Optimising the computing efficiency}
\label{sec:cmp_eff}
In order to make the code as fast as possible, several optimizations are made.
For contrast curves with a sharp inner working angle (IWA) the calculations are performed until $M_p=M_{lim}$(IWA), after which the probability map will not change.
The limits on $\rho$, $\rho_{min}$ and $\rho_{max}$ are defined by the minimum and maximum separation at which the planet is detectable given the contrast curve. At the IWA, the contrast curve increases so rapidly that effectively $\rho_{min}$ is no longer dependent on $M_p$ for $M_p>M_{lim}(IWA)$. Since $\rho_{min}$ does not change any more, the detection probability $f(M_p,a)$ no longer changes compared to that for $M_p=M_{lim}$ and thus f(M$_p>$M$_{lim}, a$)=f(M$_{lim}$, a).

If there is also an outer working angle (OWA), the code takes it into account as well.

\subsection{Comparison with the classic MESS code}
\label{sec:mess_cmp}  
In order to test the consistency of the results, 
we ran a set of simulations with identical setups, using both MESS and QMESS.
The sample used for this test is the one observed for the Gemini Deep Planet Survey \citep[GDPS][]{2007ApJ...670.1367L}, consisting of 85 young FGK stars.
In both cases we used a grid of 100 values of the $M_p$ between 1 and 75 $M_{Jup}$ and 500 values of $a$ 
between 1 and 500 AU.

\begin{figure*}[htbp]
\label{fig:map_compare}
\epsscale{0.8}
\plotone{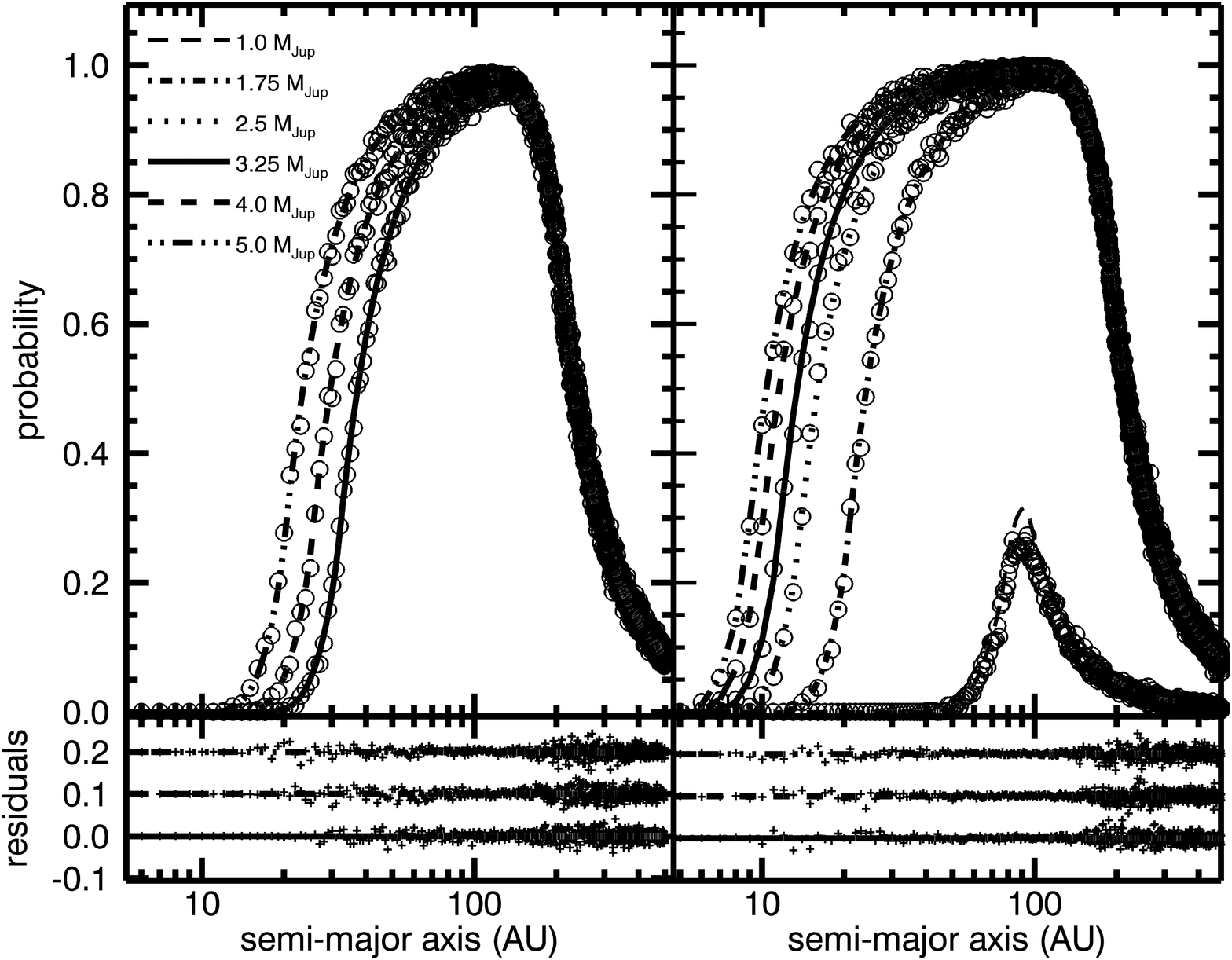}
\caption{\linespread{1} \footnotesize {\bf Top panel:} comparison of the results of MESS (open circle) and QMESS (lines) for two stars in the GDPS sample. Each line shows $f(M_p,a)$, assuming the following values of the companion mass $M_p$: 1.0, 1.75, 2.5, 3.25, 4 \& 5 $M_{Jup}$. The left panel shows only the last 3 values. {\bf Bottom panel:} residuals from the comparison between the results of MESS and QMESS, for different values of the companion mass (3.25, 4, 5 $M_{Jup}$, from bottom to top). An arbitrary offset of 0.1 has been added to help visualize the results.}
\end{figure*}

Fig.~2 shows the detection probability $f(M_p,a)$, evaluated as in Sec.~\ref{sec:detprob}, for several values of the companion mass $M_p$.
The output of the classic MESS tool (open circles) are overplotted on the ones from QMESS (lines). The same contours are shown for two stars in the GDPS sample. 
The difference between the two methods is about 5\%. This is consistent with the expected $\sim 3.2\%$ error predicted by Poisson statistic over 1000 planets per grid point generated by MESS.

\begin{figure}[htbp]
\label{fig:time}
\epsscale{0.5}
\plotone{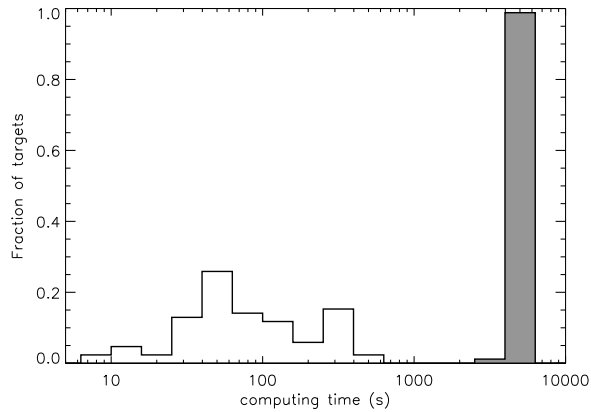}
\caption{\linespread{1} \footnotesize Fraction of the targets with a given computing time for MESS (shaded) and QMESS (unshaded) for the GDPS sample (85 stars in total). }
\end{figure} 

Fig.~3 shows the fraction of stars with a given computing time for the  85 stars in the GDPS sample, in the two test runs.
The total time needed for the full run is 2.6 hours for QMESS, as opposed to 4.2 days for the same run performed with the MESS. 
Besides the optimizations used in the QMESS code (which explains the width of the distribution of execution times, see Sec.~\ref{sec:cmp_eff}), the difference in computing time between the two codes is due to the use of the fact that the for QMESS the projection probability map is calculated once before the runs  described in Sec.~\ref{sec:prjprob}.
The QMESS therefore does not need to evaluate  the position on the projected orbit of each generated planets, which is the most time consuming step of MESS, given the number of planets involved. For the run with the GDPS sample, MESS generates 1000 planets for each step in the mass-semi-major axis grid. So a total of $4.25\cdot 10^9$ planets ($5\cdot 10^7$ per stars) are generated and need to be projected for the entire MESS run.

\section{Applications}
\label{sec:app}
 The two main purposes of the QMESS are 1) a statistical tool for the analysis of a direct imaging survey (Sec.~\ref{sec:sam}) and 2) a predictive tool (Sec.~\ref{sec:pm}) to prepare future surveys.

\subsection{Statistical Analysis of a Survey}
\label{sec:sam}
The main purpose of the QMESS is to analyse the (non-)detection from direct imaging surveys in order to test different assumptions on the planet population and to constrain the maximum occurrence of planets for a given planet population by calculating the expected number of planets and comparing that with the results from a survey with a large target sample.

As an example of this approach, we used the GDPS sample described in Sec.~\ref{sec:mess_cmp} and investigated a variety of different assumptions on the planet parameter distributions. QMESS assumes by default uniform distributions for both planetary mass and semi-major axis ($f(M_p)$ and $f(a)$). Therefore we can easily probe a wide variety of parameter distributions by folding them into the output grid. Since the properties of the planets detected via the RV \& transit methods seem to be well fitted by simple power laws \citep[e.g.][]{2008PASP..120..531C}, we explored a grid of distributions, $f(M_p)=M_p^{\alpha}$ and $f(a)=a^{\beta}$, by varying $\alpha$ and $\beta$ as well as the maximum cut-off for the semi-major axis distribution.
We used a grid of 100 steps between $M_{min}=1~M_{Jup}$ and $M_{max} =75~M_{Jup}$ for $M_p$ and a grid of 1AU steps from $a_{min}=1~AU$ to the chosen value of the cut-off for the semi-major axis.

\begin{figure}
\label{fig:abgrid}
\epsscale{1}
\plotone{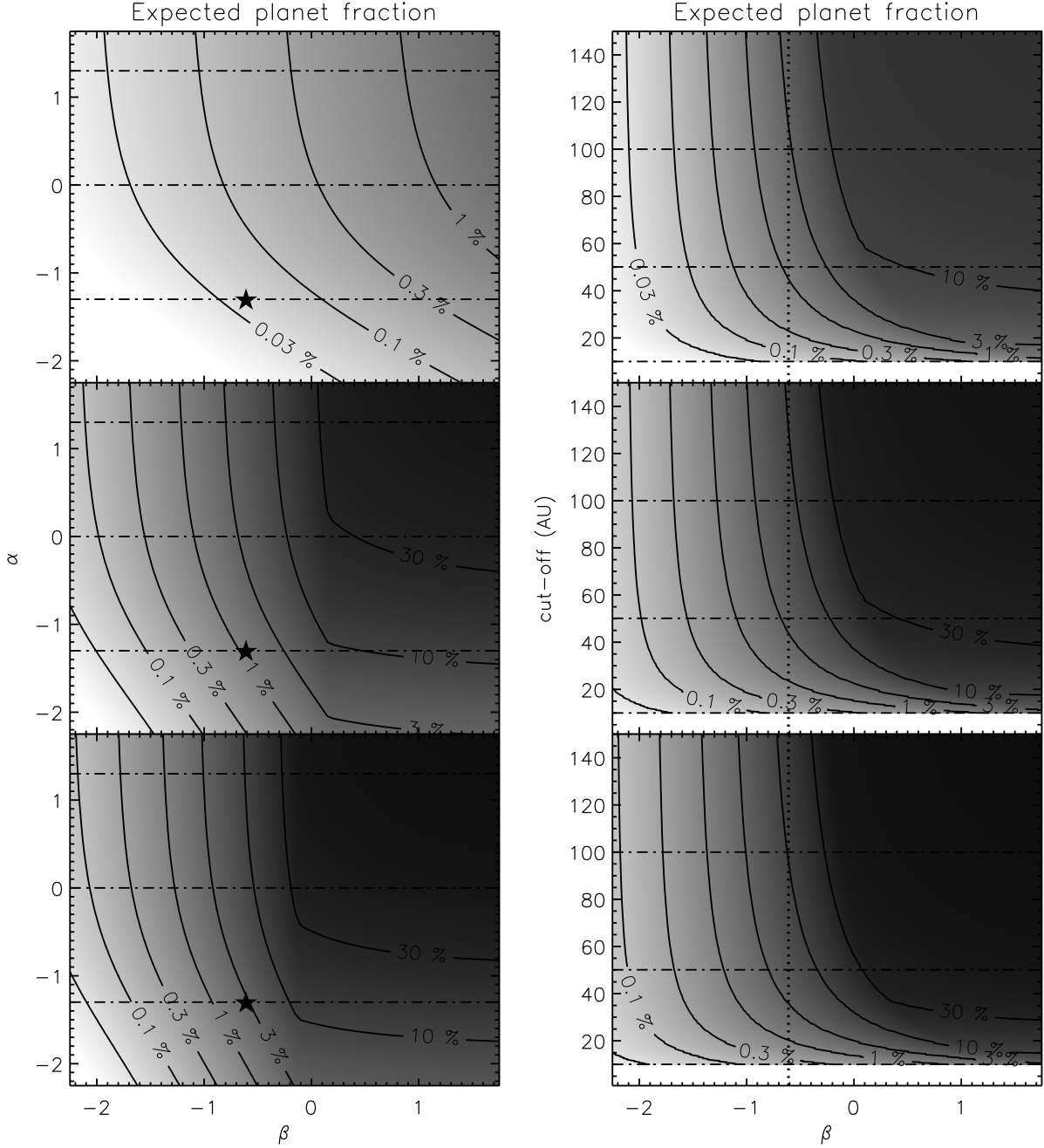}
\caption{\linespread{1} \footnotesize Expected planet fraction (evaluated using Eq.~\ref{eqn:probam}) for the GDPS sample as a function of the power law index for the mass and semi-major axis distributions (left panel) for a semi-major axis cut-off ($a_{max}$) of 10, 50 and 100 AU (bottom to top) and as a function of the power law index for the semi-major axis distribution and the semi-major axis cut-off (right panel) with a mass power-law index of -1.3, 0. \& 1.3 (bottom to top).
The mass range considered is $1 \leq M_p \leq 75$ ~$M_{Jup}$.
The values of the distributions from \cite{2008PASP..120..531C} are highlighted (star). The dotted line in the right panel corresponds to $\beta=-0.61$ \citep{2008PASP..120..531C}.  The dotted-dashed line in the left panels show the values of $\alpha$ used for the three panels on the right and in the right panels show the semi-major axis cut-off used for the three panels on the left.}
\end{figure}

The results are shown in Fig.~\ref{fig:abgrid}. The left panels show the expected planet fraction (which is equivalent to the probability $P(M_{min} \leq M_p \leq M_{max}, a_{min} \leq a \leq a_{max})$ evaluated as in Eq.~\ref{eqn:probam}) as a function of $\alpha$ and $\beta$, with the semi-major axis cut-off ($a_{max}$) fixed to 100, 50 and 10~AU (top to bottom).
The star symbol corresponds to the values reported by \cite{2008PASP..120..531C}: $\alpha=-1.31$, $\beta=-0.65$. The right panels of the same figure show the impact of varying the semi-major axis cut-off, while keeping the power law for the mass distribution fixed to $\alpha$=1.3, 0 and -1.3 (top to bottom). No planet was discovered in the GDPS survey, but \cite{2007ApJ...670.1367L} report a previously known brown dwarf companion around HD 130948, with an estimated mass in the range $40-65~M_{Jup}$ and physical separation of $\sim 47~AU$. 
Taking into account this detection, and using the same approach used by \cite{2007ApJ...670.1367L},  we estimate a frequency of brown dwarf companions ($13 M_{Jup}< M_p < 75 M_{jup}$) with separations in the range 25-225 AU of $1.4_{-1.0}^{+5.9}$ at 95\% confidence level, which is in good agreement with the value of  $1.9^{+8.3}_{-1.5}$ found by \cite{2007ApJ...670.1367L} for the same mass and semi-major axis boundaries.

For a semi-major axis cut-off of 10~AU no constraints can be made on the power-law indices. For larger cut-off values the null detection is only marginally consistent with the \cite{2008PASP..120..531C} distributions.

\subsection{Predictive Mode}
\label{sec:pm}
 In addition to the analysis of a survey, QMESS can also be used to assess the performances of, and to select the most suitable targets for new surveys, instruments and/or different observing strategies.

To demonstrate the capabilities of QMESS as a predictive tool, we evaluate the probability $P(M_{min} \leq M_p \leq M_{max}, a_{min} \leq a \leq a_{max})$ for two of the next generation planet finders for 10m-class telescopes that will soon be available to the community, The Spectro-Polarimetric High-contrast Exoplanet REsearch \citep[SPHERE, e.g.][]{2008SPIE.7014E..41B} at VLT and the Gemini Planet Finder \citep[GPI, e.g.][]{2007AAS...211.3005M} on GEMINI south.

\noindent Both instruments are expected to be available in 2013 and will target young nearby stars looking for planets in wide orbits. To assess the performances we use the estimated detection limits by \cite{2011A&A...529A.131M} for SPHERE, and \cite{2007AAS...211.3005M} for GPI. 

Whether or not it's correct to use the results of the radial velocity surveys to analyze the results of the direct imaging one it's still open to discussion. This is due to the lack of a statistically significant sample of planets in wide orbits.
This approach is although the most widely used so far for this kind of analysis, and its caveats and limitations have been discussed by several authors \citep[see e.g.][]{2012A&A...544A...9V,2010ApJ...717..878N}. For this reason, rather than investigating the impact of different distributions for mass and semi-major axis, for which we use the distributions from \cite{2008PASP..120..531C} ($f(a)=a^{-0.61}$ and $f(M_p)=M_p^{-1.3}$), we investigate the impact of two different distributions of the eccentricity  f(e): 1) uniform and 2) Gaussian ($f(e)=e^{\frac{(x-\mu)^2}{2\sigma^2}}$).  We also choose to fix the mean of the distribution, $\mu$, and its variance, $\sigma^2$, so that $\mu$=0 $\sigma$=0.3, as suggested by \cite{2010ApJ...725.2166H}. The mass range used for the simulations is the same as the one used in Sec.~\ref{sec:sam} ($M_{min}=1 ~M_{Jup}$, $M_{max}=75 ~M_{Jup}$) and we extrapolated the semi-major axis distribution from $a_{min}=1~AU$ up to $a_{max}=100~AU$.

 To explore the dependency of the planet detection probability on the characteristics of the targeted stars (e.g. spectral type, age, and distance) we generated a set of simulations for five different stellar types (A0, F0, G0, K0 \& M0), and a logarithmic grid in distance and age. Note that this simulation is not intended to represent the stellar population in the solar neighborhood, but rather to investigate the impact of differences in age, distance and stellar type on the probability of finding a planet.
The results, shown in Fig.~5, indicate that the expected planet fraction depends only weakly on the eccentricity distribution. 
The expected performances of the two instrument appear to be fairly similar and with a high ($\sim 20$~\%) chance of detection for planetary mass companions especially while targeting young, nearby stars. This is mainly due to the more favorable planet/mass contrast predicted for younger systems, and because closer in targets allow to search for companion with smaller semi-major axis (which are more likely given the assumed f(a)) at a given projected separation.

\section{Summary}
\label{sec:summary}

In this paper we describe the Quick-MESS (QMESS), a fast alternative to the classic Monte-Carlo tools for the statistical analysis of exoplanet direct imaging surveys. 
The use of a grid-based approach reduces substantially the computing times for the analysis, while reducing spurious noise from Monte Carlo sampling, as shown in Sec.~\ref{sec:mess_cmp}. 
In this paper we demonstrated the two main purposes of the QMESS and showed that, as for the MESS code, it can be used both as a statistical and predictive tool (see Sec.~\ref{sec:sam} and \ref{sec:pm}). 
For the GDPS survey we find that their null detection is marginally consistent with the distributions extrapolated from the RV results \citep{2008PASP..120..531C}. We have also shown that the eccentricity distribution only has a minor impact on the expected fraction of planet detection for the upcoming surveys, which could have a detection efficiency of up to 20\% (depending on the choice of targets) provided that the distribution of planet mass and semi-major axis are the same at large ($>1~AU$) and small ($<1~AU$) separations.

\begin{figure}
\label{fig:age_dist}
\epsscale{1.}
\plotone{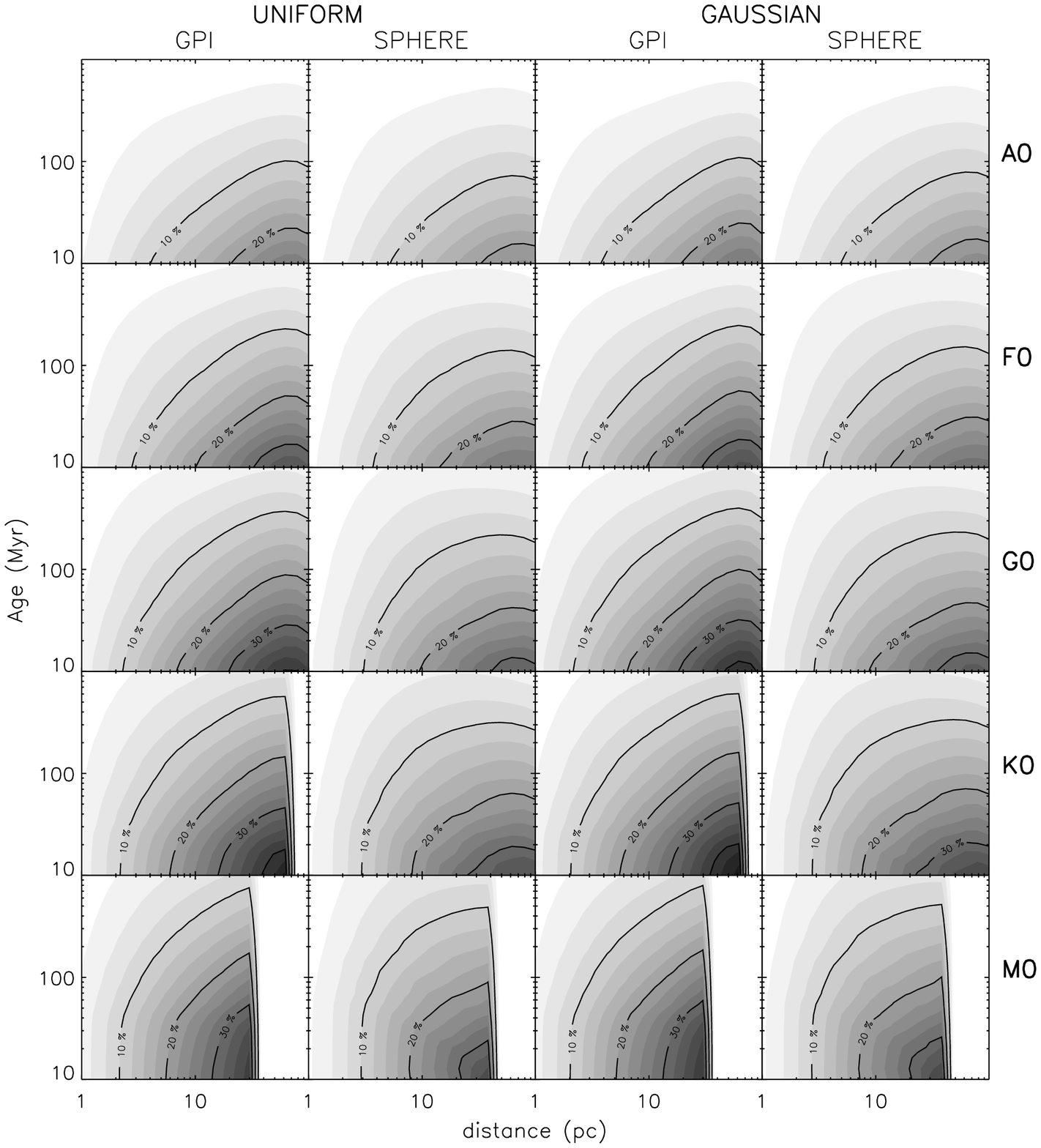}
\caption{\linespread{1} \footnotesize Comparison of the expected performances of the GPI and SPHERE integral field spectrographs \citep[see][]{2008SPIE.7014E..41B,2007AAS...211.3005M}. 
The probability $P(M_{min} \leq M_p \leq M_{max}, a_{min} \leq a \leq a_{max})$ as a function of the age, distance and spectral type for both GPI and SPHERE is shown. For the planet population  we used f(M)$\propto M^{-1.31}$, f(a)$\propto a^{-0.61}$ and f(e)=constant (left panels) and f(e)$\propto e^{\frac{-e^2}{2. 0.3^2}}$ (right). We also assumed $M_{min}=1 M_{jup}$,  $M_{max} = 75 M_{Jup}$, $a_{min}=1 AU$, $a_{max}=100 AU$ and $0 \leq e \leq 1$. The sharp cut-off for the later spectral types is due to the magnitude limits for the adaptive optic system (I$<$10 for SPHERE and I$<$9 for GPI).}
\end{figure}

\subsection*{Acknowledgments}
We thank the anonymous referee for extensive feedback that significantly improved the clarity of the paper.
E.dM. is supported in part by an Ontario Postdoctoral Fellowship. This work is supported in large part by grants to R.J. from the Natural Sciences and Engineering Research Council of Canada.


\begin{thebibliography}{21}
\expandafter\ifx\csname natexlab\endcsname\relax\def\natexlab#1{#1}\fi

\bibitem[{{Baraffe} {et~al.}(2003){Baraffe}, {Chabrier}, {Barman}, {Allard}, \&
  {Hauschildt}}]{2003A&A...402..701B}
{Baraffe}, I., {Chabrier}, G., {Barman}, T.~S., {Allard}, F., \& {Hauschildt},
  P.~H. 2003, A\&A, 402, 701

\bibitem[{{Batalha} {et~al.}(2012){Batalha}, {Rowe}, {Bryson}, {Barclay},
  {Burke}, {Caldwell}, {Christiansen}, {Mullally}, {Thompson}, {Brown},
  {Dupree}, {Fabrycky}, {Ford}, {Fortney}, {Gilliland}, {Isaacson}, {Latham},
  {Marcy}, {Quinn}, {Ragozzine}, {Shporer}, {Borucki}, {Ciardi}, {Gautier},
  {Haas}, {Jenkins}, {Koch}, {Lissauer}, {Rapin}, {Basri}, {Boss}, {Buchhave},
  {Charbonneau}, {Christensen-Dalsgaard}, {Clarke}, {Cochran}, {Demory},
  {Devore}, {Esquerdo}, {Everett}, {Fressin}, {Geary}, {Girouard}, {Gould},
  {Hall}, {Holman}, {Howard}, {Howell}, {Ibrahim}, {Kinemuchi}, {Kjeldsen},
  {Klaus}, {Li}, {Lucas}, {Morris}, {Prsa}, {Quintana}, {Sanderfer},
  {Sasselov}, {Seader}, {Smith}, {Steffen}, {Still}, {Stumpe}, {Tarter},
  {Tenenbaum}, {Torres}, {Twicken}, {Uddin}, {Van Cleve}, {Walkowicz}, \&
  {Welsh}}]{2012arXiv1202.5852B}
{Batalha}, N.~M., {Rowe}, J.~F., {Bryson}, S.~T., {et~al.} 2012, ArXiv e-prints

\bibitem[{{Beuzit} {et~al.}(2008){Beuzit}, {Feldt}, {Dohlen}, {Mouillet},
  {Puget}, {Wildi}, {Abe}, {Antichi}, {Baruffolo}, {Baudoz}, {Boccaletti},
  {Carbillet}, {Charton}, {Claudi}, {Downing}, {Fabron}, {Feautrier},
  {Fedrigo}, {Fusco}, {Gach}, {Gratton}, {Henning}, {Hubin}, {Joos}, {Kasper},
  {Langlois}, {Lenzen}, {Moutou}, {Pavlov}, {Petit}, {Pragt}, {Rabou}, {Rigal},
  {Roelfsema}, {Rousset}, {Saisse}, {Schmid}, {Stadler}, {Thalmann}, {Turatto},
  {Udry}, {Vakili}, \& {Waters}}]{2008SPIE.7014E..41B}
{Beuzit}, J.-L., {Feldt}, M., {Dohlen}, K., {et~al.} 2008, in Society of
  Photo-Optical Instrumentation Engineers (SPIE) Conference Series, Vol. 7014,
  Society of Photo-Optical Instrumentation Engineers (SPIE) Conference Series

\bibitem[{{Bonavita} {et~al.}(2012){Bonavita}, {Chauvin}, {Desidera},
  {Gratton}, {Janson}, {Beuzit}, {Kasper}, \&
  {Mordasini}}]{2012A&A...537A..67B}
{Bonavita}, M., {Chauvin}, G., {Desidera}, S., {et~al.} 2012, \aap, 537, A67

\bibitem[{{Borucki} {et~al.}(2011){Borucki}, {Koch}, {Basri}, {Batalha},
  {Brown}, {Bryson}, {Caldwell}, {Christensen-Dalsgaard}, {Cochran}, {DeVore},
  {Dunham}, {Gautier}, {Geary}, {Gilliland}, {Gould}, {Howell}, {Jenkins},
  {Latham}, {Lissauer}, {Marcy}, {Rowe}, {Sasselov}, {Boss}, {Charbonneau},
  {Ciardi}, {Doyle}, {Dupree}, {Ford}, {Fortney}, {Holman}, {Seager},
  {Steffen}, {Tarter}, {Welsh}, {Allen}, {Buchhave}, {Christiansen}, {Clarke},
  {Das}, {D{\'e}sert}, {Endl}, {Fabrycky}, {Fressin}, {Haas}, {Horch},
  {Howard}, {Isaacson}, {Kjeldsen}, {Kolodziejczak}, {Kulesa}, {Li}, {Lucas},
  {Machalek}, {McCarthy}, {MacQueen}, {Meibom}, {Miquel}, {Prsa}, {Quinn},
  {Quintana}, {Ragozzine}, {Sherry}, {Shporer}, {Tenenbaum}, {Torres},
  {Twicken}, {Van Cleve}, {Walkowicz}, {Witteborn}, \&
  {Still}}]{2011ApJ...736...19B}
{Borucki}, W.~J., {Koch}, D.~G., {Basri}, G., {et~al.} 2011, \apj, 736, 19

\bibitem[{{Burrows} {et~al.}(2003){Burrows}, {Sudarsky}, \&
  {Lunine}}]{2003ApJ...596..587B}
{Burrows}, A., {Sudarsky}, D., \& {Lunine}, J.~I. 2003, ApJ, 596, 587

\bibitem[{{Chauvin} {et~al.}(2010){Chauvin}, {Lagrange}, {Bonavita},
  {Zuckerman}, {Dumas}, {Bessell}, {Beuzit}, {Bonnefoy}, {Desidera}, {Farihi},
  {Lowrance}, {Mouillet}, \& {Song}}]{2010A&A...509A..52C}
{Chauvin}, G., {Lagrange}, A.-M., {Bonavita}, M., {et~al.} 2010, \aap, 509, A52

\bibitem[{{Cumming} {et~al.}(2008){Cumming}, {Butler}, {Marcy}, {Vogt},
  {Wright}, \& {Fischer}}]{2008PASP..120..531C}
{Cumming}, A., {Butler}, R.~P., {Marcy}, G.~W., {et~al.} 2008, PASP, 120, 531

\bibitem[{{Fischer} \& {Valenti}(2005)}]{2005ApJ...622.1102F}
{Fischer}, D.~A. \& {Valenti}, J. 2005, ApJ, 622, 1102

\bibitem[{{Heintz}(1978)}]{1978GAM....15.....H}
{Heintz}, W.~D. 1978, Geophysics and Astrophysics Monographs, 15

\bibitem[{{Hogg} {et~al.}(2010){Hogg}, {Myers}, \&
  {Bovy}}]{2010ApJ...725.2166H}
{Hogg}, D.~W., {Myers}, A.~D., \& {Bovy}, J. 2010, \apj, 725, 2166

\bibitem[{{Johnson} {et~al.}(2007){Johnson}, {Butler}, {Marcy}, {Fischer},
  {Vogt}, {Wright}, \& {Peek}}]{2007ApJ...670..833J}
{Johnson}, J.~A., {Butler}, R.~P., {Marcy}, G.~W., {et~al.} 2007, ApJ, 670, 833

\bibitem[{{Lafreni{\`e}re} {et~al.}(2007){Lafreni{\`e}re}, {Doyon}, {Marois},
  {Nadeau}, {Oppenheimer}, {Roche}, {Rigaut}, {Graham}, {Jayawardhana},
  {Johnstone}, {Kalas}, {Macintosh}, \& {Racine}}]{2007ApJ...670.1367L}
{Lafreni{\`e}re}, D., {Doyon}, R., {Marois}, C., {et~al.} 2007, ApJ, 670, 1367

\bibitem[{{Lafreni{\`e}re} {et~al.}(2008){Lafreni{\`e}re}, {Jayawardhana}, \&
  {van Kerkwijk}}]{2008ApJ...689L.153L}
{Lafreni{\`e}re}, D., {Jayawardhana}, R., \& {van Kerkwijk}, M.~H. 2008, \apjl,
  689, L153

\bibitem[{{Lineweaver} \& {Grether}(2003)}]{2003ApJ...598.1350L}
{Lineweaver}, C.~H. \& {Grether}, D. 2003, ApJ, 598, 1350

\bibitem[{{Macintosh} {et~al.}(2007){Macintosh}, {Graham}, {Palmer}, {Doyon},
  {Larkin}, {Oppenheimer}, {Saddlemyer}, {Veran}, {Wallace}, \& {Gemini Planet
  Imager team}}]{2007AAS...211.3005M}
{Macintosh}, B., {Graham}, J.~R., {Palmer}, D., {et~al.} 2007, in Bulletin of
  the American Astronomical Society, Vol.~38, Bulletin of the American
  Astronomical Society, 782--+

\bibitem[{{Mesa} {et~al.}(2011){Mesa}, {Gratton}, {Berton}, {Antichi},
  {Verinaud}, {Boccaletti}, {Kasper}, {Claudi}, {Desidera}, {Giro}, {Beuzit},
  {Dohlen}, {Feldt}, {Mouillet}, {Chauvin}, \& {Vigan}}]{2011A&A...529A.131M}
{Mesa}, D., {Gratton}, R., {Berton}, A., {et~al.} 2011, \aap, 529, A131

\bibitem[{{Nielsen} \& {Close}(2010)}]{2010ApJ...717..878N}
{Nielsen}, E.~L. \& {Close}, L.~M. 2010, \apj, 717, 878

\bibitem[{{Nielsen} {et~al.}(2008){Nielsen}, {Close}, {Biller}, {Masciadri}, \&
  {Lenzen}}]{2008ApJ...674..466N}
{Nielsen}, E.~L., {Close}, L.~M., {Biller}, B.~A., {Masciadri}, E., \&
  {Lenzen}, R. 2008, ApJ, 674, 466

\bibitem[{{Santos} {et~al.}(2004){Santos}, {Israelian}, \&
  {Mayor}}]{2004A&A...415.1153S}
{Santos}, N.~C., {Israelian}, G., \& {Mayor}, M. 2004, A\&A, 415, 1153

\bibitem[{{Vigan} {et~al.}(2012){Vigan}, {Patience}, {Marois}, {Bonavita}, {De
  Rosa}, {Macintosh}, {Song}, {Doyon}, {Zuckerman}, {Lafreni{\`e}re}, \&
  {Barman}}]{2012A&A...544A...9V}
{Vigan}, A., {Patience}, J., {Marois}, C., {et~al.} 2012, \aap, 544, A9

\end{thebibliography}

\end{document}